\documentclass[final]{aipproc} 

\layoutstyle{8x11single}

\usepackage[matrix,frame,arrow]{xy}
\usepackage{amsmath}
\newcommand{\bra}[1]{\left\langle{#1}\right\vert}
\newcommand{\ket}[1]{\left\vert{#1}\right\rangle}
\newcommand{\qw}[1][-1]{\ar @{-} [0,#1]}
\newcommand{\qwx}[1][-1]{\ar @{-} [#1,0]}
\newcommand{\cw}[1][-1]{\ar @{=} [0,#1]}
\newcommand{\cwx}[1][-1]{\ar @{=} [#1,0]}

\newcommand{\measurebub}[1]{*+[F-:<.9em>]{#1} \qw}

\newcommand{\control}{*-=-{\bullet}}
\newcommand{\controlo}{*-=[o][F]{\phantom{\bullet}}}
\newcommand{\ctrl}[1]{\control \qwx[#1] \qw}
\newcommand{\ctrlo}[1]{\controlo \qwx[#1] \qw}
\newcommand{\targ}{*{\xy{<0em,0em>*{} \ar @{ - } +<.4em,0em> \ar @{ - } -<.4em,0em> \ar @{ - } +<0em,.4em> \ar @{ - } -<0em,.4em>},*+<.8em>\frm{o}\endxy} \qw}

\newcommand{\multigate}[2]{*+{\hphantom{#2}} \qw \POS[0,0].[#1,0] !C *{#2} \POS[0,0].[#1,0] \drop\frm{-}}
\newcommand{\ghost}[1]{*+{\hphantom{#1}} \qw}

\newcommand{\lstick}[1]{*!R!<.5em,0em>=<0em>{#1}}

\newcommand{\Qcircuit}{\xymatrix @*=<0em>}

\def\ketc[#1]{\vert #1 \rangle}
\def\brac[#1]{\langle #1 \vert}

\newcommand{\nn}{\nonumber}
\newcommand{\dg}{^\dagger}

\begin{document}

\title{Continuous quantum error correction}

\author{Mohan Sarovar}{
address={Centre for Quantum
Computer Technology, and School of Physical Sciences, The
University of Queensland, St Lucia, QLD 4072, Australia}
}

\author{G. J. Milburn}{
address={Centre for Quantum
Computer Technology, and School of Physical Sciences, The
University of Queensland, St Lucia, QLD 4072, Australia}
}


\begin{abstract}
We describe new implementations of quantum error correction that are continuous in time, and thus described by continuous dynamical maps. We evaluate the performance of such schemes using numerical simulations, and comment on the effectiveness and applicability of continuous error correction for quantum computing. 
\end{abstract}

\maketitle


\section{Introduction}
\label{sec:intro}

Error correction is an essential ingredient for quantum computing and communication. The combined use of error correction and fault tolerance techniques allows a quantum computation to succeed even in the presence of errors during the computation. Quantum error correction is possible because of the ability of quantum error correcting codes (QECC) to redundantly encode quantum information and permit its retrieval. The first of these codes were discovered by Shor \cite{shor95} and Steane \cite{steane96}, and these discoveries sparked a slew of work in quantum error correcting codes, part of which has resulted in the elegant theory of stabilizer codes \cite{gottesman}. Stabilizer codes are the best known (class of) QECC to date and most error correction protocols involve their use \footnote{For completeness, we mention that  there are also other techniques that protect quantum information against errors (e.g. decoherence free subsystems \cite{dfs}).}. Error correction schemes that involve the use of QECC and the detection and correction operations that go along with them are referred to as \textit{active error correction} schemes.

\begin{figure}
\includegraphics[scale=0.35]{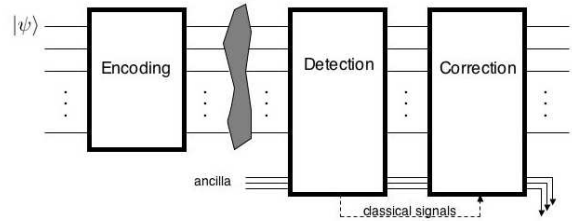}
\caption{Standard active error correction implementation blueprint. The communication between the detection and correction blocks will require classical signals for error correction with measurement.} \label{fig:aec_gen}
\end{figure}

Figure \ref{fig:aec_gen} illustrates the standard blueprint for implementing active error correction schemes. The procedure begins by encoding of the (unknown) qubit state to be protected using an error correcting code. Then after a period of time during which errors could occur, an error detection operation is performed which, assuming the code used can handle the number and type of errors that has occurred, yields information about error type and location. This information is used to quantum mechanically or classically condition the next block - the correction - which performs operations on the encoded block of qubits to reverse the effect of the error. If the conditioning between the detect and correct blocks is quantum mechanical, then this is referred to as error correction \textit{without measurement}. If on the other hand, the conditioning is classical then it is referred to as error correction \textit{with measurement}, and requires classical communication between the two blocks. 

The detection and correction operations are often repeated many times, and the whole sequence concludes with a decoding operation that retrieves the protected one qubit state from the encoded block. A feature of this blueprint we would like to emphasize is its discrete nature. The detection and correction operations occur at discrete points in time and they are usually assumed to be instantaneous and arbitrarily accurate. For this reason we will refer to such an implementation model as \textit{discrete error correction}. In this paper we will present an alternative to this model where both the detection and correction blocks operate continuously and occur at a certain rate as opposed to instantaneously. This will lead to a description of error correction in terms of continuous dynamical maps, which we will solve numerically to evaluate error correcting capabilities.

There have been a number of authors in the past who have considered continuous forms of error correction. Paz and Zurek \cite{pazzurek} examined the limit where the detection and correction operations are performed directly after each other continually. However, this is not quite continuous error correction in the sense of this paper because they still assumed both operations occur instantaneously. Mabuchi and Zoller \cite{mabuchizoller}, and Ahn \textit{et. al.} \cite{ahn-wm} have examined continuous error correction for a specific subset of error models (detected errors). Mabuchi and Zoller also assume that the correction operations are instantaneous, and therefore treat a slightly different case from us. Finally, Ahn \textit{et. al} in \cite{ahn-dl} examine the exact case we will cover in the section entitled \textit{continuous error correction by indirect feedback}. However, the error correction scheme they describe involves intensive real-time computation, and is therefore considerably less practical than the scheme we detail.

\section{Review of concepts}
\label{sec:review}

\subsection{Quantum feedback control}
Making the detection and correction operations continuous places the process of error correction within the framework of quantum feedback control \cite{Belavkin83, wiseman-fb, wiseman-thesis, Habib-JM, DHJMT00} -  a field which examines the control of quantum systems using closed loop control methods. We will use the tools of this field to describe the error correction procedures below. 

We will not cover the field of quantum feedback control in any detail here, for that the reader is referred to \cite{DHJMT00, Habib-JM} and references therein. However, it is useful for what follows to make a distinction between the two types of quantum feedback control. This distinction arises essentially because the controller steering the quantum system of interest can itself be either quantum or classical. The two alternatives lead to different control loops and hence different mathematical descriptions. This distinction is also presented in \cite{Habib-JM} and we shall use the terminology used there. 

A \textit{direct feedback} control loop has a quantum controller that is directly coupled to variables of interest in the quantum system to be controlled. The interaction between the system and controller in this case is unitary. In an \textit{indirect feedback} loop, measurements are performed on a quantum system and the results of these measurements are used by a classical controller to form classical conditioning signals to control the system. 

\subsection{The bit-flip code}
As mentioned earlier, stabilizer codes are the best known QECCs. We will not attempt a survey of the extensive theory of stabilizer codes (the interested reader is referred to \cite{gottesman, mikeandike}), but will rather describe one of the simplest stabilizer codes - the bit-flip code - which will be used in the following sections to illustrate the continuous error correction schemes.  

The bit-flip code is essentially a classical parity check code. It protects against a logical flip error which takes one computational basis state to the other. The bit-flip code protects against this error by using the following repetition encoding: $\ket{0}_L \equiv \ket{000}_P, ~ \ket{1}_L \equiv \ket{111}_P$, where the subscripts L and P stand for logical and physical, respectively. Therefore a general encoded qubit will have the form $\ket{\psi} = \alpha\ket{0}_L + \beta\ket{1}_L = \alpha\ket{000} + \beta\ket{111}$. We will refer to the encoded qubit states as the \textit{codewords}, and the subspace they span as the \textit{codespace}.

This code can detect and correct one bit-flip. The detection operation involves measuring the operators $ZZI$ and $IZZ$ \footnote{We denote the Pauli $\sigma_X, \sigma_Y, \textrm{and } \sigma_Z$ operators by $X$, $Y$, and $Z$, respectively, and suppress the tensor product sign. Therefore $ZZI \equiv Z\otimes Z \otimes I$.}, which are referred to as the \textit{error syndromes} or \textit{stabilizer generators}. Two things to note, both of which are properties of general stabilizer codes, are that the stabilizer generators commute, and that the codewords are both eigenvalue one eigenstates of the stabilizer generators (or in other words, the codespace is stabilized by the generators).

The four possible outcomes from measuring the two stabilizer generators tell us about the four possible error events. This is illustrated by table \ref{tab:tq_table}. Correcting errors using this code then simply amounts to applying a unitary to restore the encoded state back to its unperturbed value. The value of this unitary depends on the measurement results as table \ref{tab:tq_table} shows.

\begin{table}
\begin{tabular}{|c|c|c|c|}
  \hline
  \textbf{$\langle ZZI\rangle_\rho$} & \textbf{$\langle IZZ\rangle_\rho$} & \textbf{Error} & \textbf{Correcting unitary} \\
  \hline\hline
  +1 & +1 & None & None \\
  -1 & +1 & on qubit 1 & XII \\
  +1 & -1 & on qubit 3 & IIX \\
  -1 & -1 & on qubit 2 & IXI \\
  \hline
\end{tabular}
\caption{The three qubit bit-flip code. Note that each error results in a different sequence of stabilizer generator measurement results. $\langle \cdot \rangle_\rho$ represents the expectation value of $\cdot$ under the encoded three qubit state $\rho$.} \label{tab:tq_table}
\end{table}

\section{Continuous error correction by indirect feedback}
\label{sec:indirect}

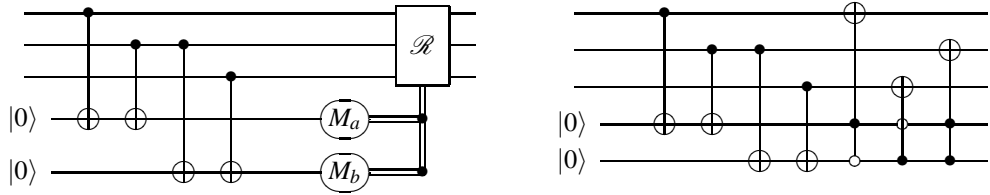
\begin{figure}
\begin{array}[t]{ccc}
\Qcircuit @C=1em @R=.6em 
{ & \qw & \ctrl{3} & \qw & \qw & \qw & \qw & \qw & \qw & \multigate{2}{\ \mathcal{R}\ } & \qw\\ 
& \qw & \qw & \ctrl{2} & \ctrl{3} & \qw & \qw & \qw & \qw & \ghost{\ \mathcal{R}\ } \qw & \qw\\ 
& \qw & \qw & \qw & \qw & \ctrl{2} & \qw & \qw & \qw & \ghost{\ \mathcal{R}\ } \qw & \qw\\ 
& \lstick{\ket{0}} & \targ \qw & \targ \qw & \qw & \qw & \qw & \qw & \measurebub{M_a} & \control \cw \cwx\\ 
& \lstick{\ket{0}} & \qw & \qw & \targ \qw & \targ \qw & \qw & \qw & \measurebub{M_b} & \control \cw \cwx\\} &~~~~~~~&
\Qcircuit @C=1em @R=0.6em { 
& \qw & \qw & \ctrl{3} & \qw      & \qw      & \qw      & \targ        & \qw          & \qw       & \qw \\
& \qw & \qw & \qw      & \ctrl{2} & \ctrl{3}  \qw   & \qw & \qw          & \qw          & \targ     & \qw \\
& \qw & \qw & \qw      & \qw      & \qw & \ctrl{2}      & \qw          & \targ        & \qw       & \qw \\
& \lstick{\ket{0}} & \qw &  \targ   & \targ     &   \qw  & \qw      & \ctrl{-3}   & \ctrlo{-1} & \ctrl{-2} & \qw \\
& \lstick{\ket{0}} & \qw &  \qw     &  \qw     & \targ     & \targ    & \ctrlo{-4} & \ctrl{-2}   & \ctrl{-3} & \qw } \end{array}
\caption{Circuits for implementing the three qubit bit-flip code. The circuit on the left shows an implementation that uses measurement, and the circuit on the right shows one that uses no measurement. In both figures, the top three qubits form the encoded logical qubit and the bottom two are ancilla. Note that for the circuit on the right, to repeat the error correction procedure, the ancilla qubits must be replaced or reset to the $\ket{0}$ state at the end of each run (at the far right of the circuit).} \label{fig:circuits}
\end{figure}

The indirect feedback model of error correction is a continuous version of discrete error correction \textit{with measurement}. Full details on this model are in the paper \cite{SAJM04}, we will simply outline the results. The subfigure on the left of figure \ref{fig:circuits} shows a circuit which implements the bit-flip code in a discrete manner with measurement. The top three qubits in the circuit represent the encoded qubit and the bottom two represent ancilla qubits used in the stabilizer generator measurements. The first part of the circuit, the CNOT gates and the projective measurements, detect the error by placing the results of the stabilizer generator measurements in the ancilla qubits and measuring in the computational basis. Then the results of these measurements are used to condition the unitary correcting gate, $\mathcal{R}$. To make this process continuous we must replace the CNOT gates and projective measurements by a weak, continuous measurement of the stabilizer generators, and the unitary correcting gate by Hamiltonian evolution which is conditioned on the continuous stabilizer measurement record. 

Note that for the three qubit code we need to measure two stabilizer generators continuously and simultaneously. This poses no fundamental problems because by definition the stabilizer generators are commuting observables. Also, because the codespace is stabilized by the generators, when there is no error, these measurements do not affect the encoded qubits at all. However, one problem that does arise is that the continuous measurements, because they are weak, are very noisy. This is just a manifestation of the tradeoff in quantum mechanics between the strength of a measurement (and hence the amount of back action on the system being measured) and the the amount of information gained from the measurement. As in classical feedback control, feedback conditioned on noisy measurement records is in general ineffective, and thus we cannot perform error correction by using the raw measurement signals. Instead we must insert a signal processing step between the measurement and the classical controller which smoothes the stabilizer generator measurement records. 

Therefore the steps involved in the error correcting scheme we propose are:
\begin{enumerate}
\item Encode information in a stabilizer code suited to the errors of concern. \item Continuously perform weak measurements of the stabilizer generators, and smooth the measurement records. \item From these smoothed measurement records, form conditioning signals for feedback operators on each physical qubit. The mapping between measurement results and conditioning signals is determined by the stabilizer code. \item Apply feedback (correction) Hamiltonians to each physical qubit, where the strength of the Hamiltonians is given by the conditioning signals formed in the previous step.
\end{enumerate}

For the bit-flip code, weak measurements are performed of the observables $ZZI$ and $IZZ$. The feedback Hamiltonians are $XII$, $IXI$, and $IIX$, each one implementing a correction on one of the physical qubits in the encoding. The mapping between the results of the (smoothed) measurements and the feedback conditioning signals is essentially a translation of table \ref{tab:tq_table}.

We use the language of open quantum systems \cite{gardiner, quantumnoise, breuer_oqs} to model this process. The following \textit{stochastic master equation} (SME) describes the evolution of the encoded three qubit system while undergoing random bit-flip errors, weak measurement of the stabilizer generators, and continuous feedback conditioned on the measurement results \cite{SAJM04}.
\begin{eqnarray}
\label{eq:tq_drho_fb} 
d\rho_c(t) &=& \gamma(\mathcal{D}[XII]+\mathcal{D}[IXI]+\mathcal{D}[IIX])\rho_c(t)
dt + \kappa(\mathcal{D}[ZZI] + \mathcal{D}[IZZ])\rho_c(t) dt  
+ \sqrt{\kappa}(\mathcal{H}[ZZI]dW_1(t) \nn \\ &+& \mathcal{H}[IZZ]dW_2(t))\rho_c(t)  - i\lambda(G_1(t)[XII, \rho_c(t)] + G_2(t)[IXI, \rho_c(t)] + G_3(t)[IIX, \rho_c(t)])dt
\end{eqnarray}
where $\gamma$ is the error rate for each qubit, $\kappa$ is the measurement strength, and $\lambda$ is the maximum feedback strength. $G_i(t)$ are the feedback conditioning signals formed from the smoothed measurement records, the quantities $dW_i(t)$ are \textit{Weiner increments} \cite{gardiner}, and $\mathcal{H}$ and $\mathcal{D}$ are the superoperators
\begin{equation}
\label{eq:supop}
\mathcal{D}[A]\rho = A\rho A\dg - \frac{1}{2}(A\dg A\rho + \rho A\dg A), ~~~~~~
\mathcal{H}[A]\rho = A\rho + \rho A\dg - \rho ~\text{tr}[A\rho + \rho A\dg]
\end{equation}
for any operator $A$. Also note that we set $\hbar=1$ throughout this paper.

Eq. \eqref{eq:tq_drho_fb} is a non-Markovian, non-linear SME, and therefore the only way to solve it is through numerical simulation. We did precisely this (for details see \cite{SAJM04}), to evaluate the efficacy of this error correction scheme. The results of these simulations are summarized by the left subfigure of figure \ref{fig:fids}. We used fidelity with the initial state, $F(t) = \bra{\psi_0}\rho(t)\ket{\psi_0}$ as a figure of merit to evaluate the error correction scheme. The figure shows fidelity versus time curves for several values of error rate ($\gamma$).  Each plot also shows the fidelity curve for one qubit in the absence of error correction. A comparison of these two curves shows that the fidelity is preserved for a longer period of time by the error correction scheme for small enough error rates. Furthermore, for small error rates ($\gamma<0.3$) the solid curve shows a vast improvement over the exponential decay in the absence of error correction. However, we see that as the error rate increases, the fidelity decay even in the presence of error correction behaves exponentially, and the two curves look very similar. When the error rate gets large enough, the error correcting scheme becomes unable to handle the errors and becomes ineffective.  

\section{Continuous error correction by direct feedback}
\label{sec:direct}

The direct feedback model of error correction is a continuous version of discrete error correction \textit{without measurement}. To illustrate this model we will again describe the continuous analogue of a circuit that implements the bit-flip code: the subfigure on the right of figure \ref{fig:circuits}. In this circuit the results of the stabilizer generators are again placed in the ancilla qubits, however, instead of measurements followed by classical conditioning of the correction operation, the correction is done by direct coupling between the ancilla and the encoded qubits (via Toffoli gates in this case). It is important to note that the ancilla qubits must be reset to the $\ket{0}$ state after each run of the circuit. This is essentially a consequence of the no-deleting theorem \cite{nodel} for quantum information and because the error is moved from the encoded qubits into the ancilla array.

The process of making this circuit continuous is much simpler than in the indirect feedback case. We simply have to model the detection and correction unitary gates as one continuous Hamiltonian evolution. However in addition to this, we must also replace the ancilla reset procedure with a continuous map. The natural candidate is a \textit{cooling} process that continuously damps each ancilla independently. Both these processes, along the with random bit-flip errors, can be modeled by the following simple master equation (ME):
\begin{equation}
\label{eq:direct_me}
\frac{d\rho}{dt} = \gamma(\mathcal{D}[XIIII] + \mathcal{D}[IXIII] + \mathcal{D}[IIXII])\rho + \lambda(\mathcal{D}[IIIS^-I] + \mathcal{D}[IIIIS^-])\rho -i\kappa[H, \rho]
\end{equation}
where $\gamma$ is the bit-flip error rate, $\kappa$ is the strength of $H$, the Hamiltonian evolution which performs the detection and correction, and $\lambda$ is the rate of the cooling applied to the ancilla qubits. The super-operator $\mathcal{D}$ is defined as in Eq. \eqref{eq:supop}, and $S^- \equiv \frac{1}{2}(X+iY) = \ket{0}\bra{1}$ is the qubit lowering operator. In the above equation, the ordering of the tensor product runs down the circuit (i.e. the first three operators apply to the encoded qubit, and the last two to the ancilla).

\begin{figure}
\begin{array}[t]{ccc}
\includegraphics[scale=0.4]{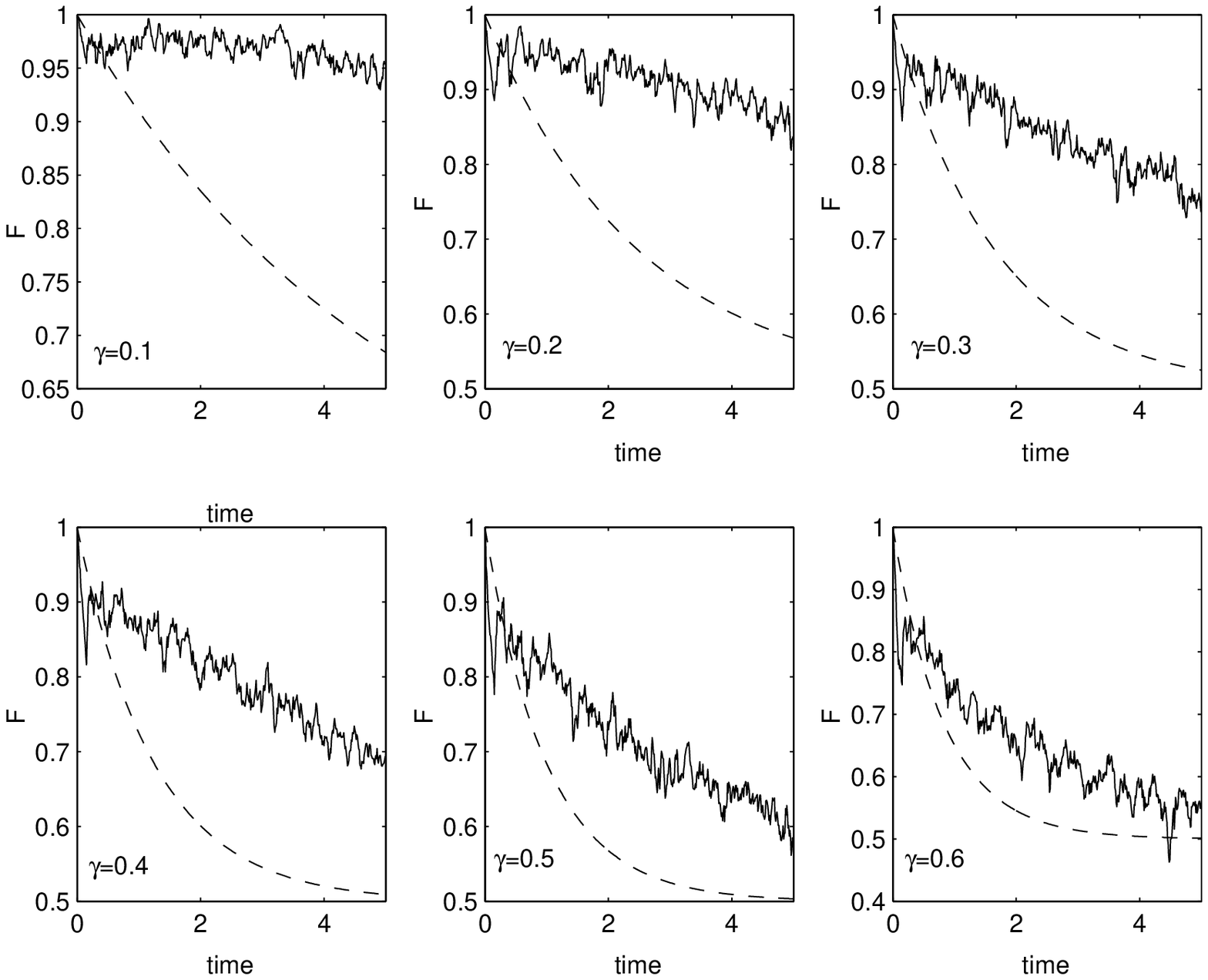}
& ~~~~ & \includegraphics[scale=0.4]{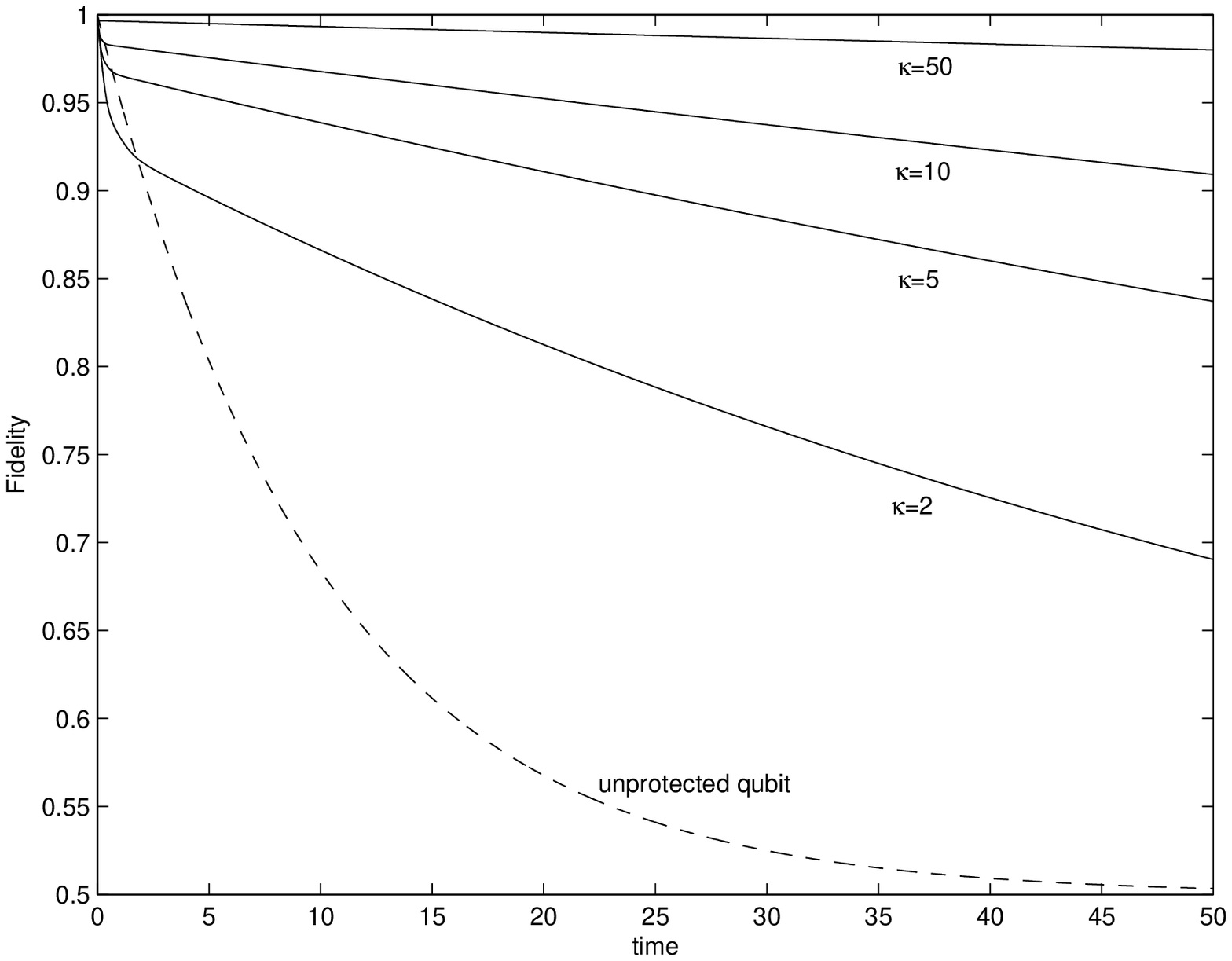}
\end{array}
\caption{Fidelity curves for the two error correcting schemes. The figure on the left shows curves for several error rates for the \textit{indirect feedback} model. The solid curves are the fidelities of the encoded qubit (three qubit code) with parameters: $\kappa=150~Hz, \lambda=150~Hz$, initial state $\ket{\psi_0} = \ket{000}$). The figure on the right shows curves for several Hamiltonian strength and cooling rates for the \textit{direct feedback} model. The solid curves are the fidelities of an encoded qubit (three qubit code) with parameters: $\gamma=0.05~Hz, \lambda=2.5\times\kappa$, initial state $\ket{\psi_0} = \ket{000}$. The dashed curves on both figures show the fidelity of one qubit undergoing random bit-flips without error correction (initial state $\ket{\psi_0} = \ket{0}$).} 
\label{fig:fids}
\end{figure}

As before, we can evaluate the efficacy of this scheme by numerically solving the ME \eqref{eq:direct_me}. The results of this are summarized by the right subfigure of figure \ref{fig:fids}, where again, we have used fidelity with the initial state as the figure of merit. More details on this scheme and the numerical solutions are in \cite{SM04}.

We see from figure \ref{fig:fids} that the ability of the direct feedback scheme to preserve fidelity improves as the Hamiltonian strength ($\kappa$) and cooling rate ($\lambda$) are increased. This is to be expected because as these parameters are increased, the whole scheme approaches a discrete error correction process (which is the $\kappa=\lambda \rightarrow \infty$ limit) operating continually. However, it is important to note that we cannot improve performance by simply increasing one parameter; the other must also be  proportionally increased.

\section{Assessment and conclusion}
\label{sec:conc}
The aim of this paper is to show that error correction may be effectively performed using continuous (in time) implementations. We have described two continuous implementations that are direct analogies of standard discrete implementations, and validated their efficacy through numerical simulations. Two concluding points about the schemes presented are: (1) Even though the schemes were evaluated for the bit-flip code, it easy to see how one would extend them for any stabilizer code; and (2) Both schemes have a number of free parameters that can be optimized over. We expect the performance of such optimized versions to be noticeably better than the results presented here.

At this point, we can make the following assessment on the advantages and disadvantages of continuous error correction implementations:
\begin{itemize}
\item Advantages
\begin{enumerate}
\item Such models provide a useful framework within which to approach quantum error correction (QEC) from a quantum control framework.
\item Continuous implementations of QEC will be more suited to most quantum computing architectures - at least in the short term - because the projective measurements and/or fast unitary gates needed for discrete time implementations are not available in most architectures.
\item Both models can be seen as providing an upper bound for the performance of an error correcting code when measurements, gates, and resetting operations are not instantaneous. This is because they simulate the situation where such non-instantaneous operations are \textit{continuously} applied to combat errors - this is an ideal scenario because in reality these operations would be interspersed with actual computation operations. 
\end{enumerate}
\item Disadvantages
\begin{enumerate}
\item Continuous implementations seem ideal for preserving quantum memory, but are less suited to computation because it is not clear that the computation operations can be performed while the error correction procedure is on. 
\item In the indirect feedback model, as the QECC being implemented becomes larger, the measurement and feedback rates must increase for the scheme to stay effective. This can be a problem practically because these rates could be limited by physical restrictions. 
\item In the direct feedback model, the entire set of unitary operations that correspond to the detection and correction operations need to be implemented as a Hamiltonian evolution. Depending on the physical system, setting up such a Hamiltonian can be a difficult thing to do. 
\end{enumerate}
\end{itemize}

There are a number of interesting directions in which to continue the investigation of continuous error correction implementations. The first of these is to identify good parameter regimes for both of the schemes described above. This is important because of the strong dependency of the performance of both schemes on a good choice of parameters. Another direction is to investigate how fault tolerance is affected by continuous implementations. Similarly, such implementation models could be useful in answering the question: how does the fault tolerance threshold change when the operations implementing error correction are not assumed to be instantaneous and arbitrarily accurate? 

\begin{theacknowledgments}
We gratefully acknowledge the support of the Australian Research Council Centre of Excellence in Quantum Computer Technology.
\end{theacknowledgments}
 
\bibliographystyle{aipproc} 
\bibliography{qcmc_proc}

\begin{thebibliography}{20}
\expandafter\ifx\csname natexlab\endcsname\relax\def\natexlab#1{#1}\fi
\providecommand{\enquote}[1]{``#1''}
\expandafter\ifx\csname url\endcsname\relax
  \def\url#1{\texttt{#1}}\fi
\expandafter\ifx\csname urlprefix\endcsname\relax\def\urlprefix{URL }\fi

\bibitem[Shor(1995)]{shor95}
Shor, P., \emph{Phys. Rev. A}, \textbf{52}, 2493 (1995).

\bibitem[Steane(1996)]{steane96}
Steane, A.~M., \emph{Phys. Rev. Lett.}, \textbf{77}, 793 (1996).

\bibitem[Gottesman(1997)]{gottesman}
Gottesman, D., \emph{Stabilizer codes and quantum error correction}, Ph.D.
  thesis, Caltech (1997).

\bibitem[Lidar and Whaley(2003)]{dfs}
Lidar, D.~A., and Whaley, K.~B., \emph{quant-ph/0301032} (2003).

\bibitem[Paz and Zurek(1997)]{pazzurek}
Paz, J.~P., and Zurek, W.~H., \emph{quant-ph/9707049} (1997).

\bibitem[Mabuchi and Zoller(1996)]{mabuchizoller}
Mabuchi, H., and Zoller, P., \emph{Phys. Rev. Lett.}, \textbf{76}, 3108 (1996).

\bibitem[Ahn et~al.(2003)]{ahn-wm}
Ahn, C., Wiseman, H.~M., and Milburn, G.~J., \emph{Phys. Rev. A}, \textbf{67},
  052310 (2003).

\bibitem[Ahn et~al.(2002)]{ahn-dl}
Ahn, C., Doherty, A.~C., and Landahl, A.~J., \emph{Phys. Rev. A}, \textbf{65},
  042301 (2002).

\bibitem[Belavkin(1983)]{Belavkin83}
Belavkin, V.~P., \emph{Automatica and Remote Control}, \textbf{44 (2)}, 178
  (1983).

\bibitem[Wiseman(1994{\natexlab{a}})]{wiseman-fb}
Wiseman, H.~M., \emph{Phys. Rev. A}, \textbf{49}, 2133 (1994{\natexlab{a}}).

\bibitem[Wiseman(1994{\natexlab{b}})]{wiseman-thesis}
Wiseman, H.~M., \emph{Quantum Trajectories and Feedback}, Ph.D. thesis, The
  University of Queensland (1994{\natexlab{b}}).

\bibitem[Habib et~al.(2002)]{Habib-JM}
Habib, S., Jacobs, K., and H., M., \emph{Los Alamos Science}, \textbf{27}, 126
  (2002).

\bibitem[Doherty et~al.(2000)]{DHJMT00}
Doherty, A.~C., Habib, S., Jacobs, K., Mabuchi, H., and Tan, S.~M., \emph{Phys.
  Rev. A}, \textbf{62}, 012105 (2000).

\bibitem[Nielsen and Chuang(2000)]{mikeandike}
Nielsen, M.~A., and Chuang, I.~L., \emph{Quantum Computation and Quantum
  Information}, Cambridge University Press, 2000.

\bibitem[Sarovar et~al.(2004)]{SAJM04}
Sarovar, M., Ahn, C., Jacobs, K., and Milburn, G.~J., \emph{Phys. Rev. A},
  \textbf{69}, 052324 (2004).

\bibitem[Gardiner(1985)]{gardiner}
Gardiner, C.~W., \emph{Handbook of Stochastic Methods}, Springer, 1985.

\bibitem[Gardiner and Zoller(2000)]{quantumnoise}
Gardiner, C.~W., and Zoller, P., \emph{Quantum Noise}, Springer, 2000.

\bibitem[Breuer and Petruccione(2002)]{breuer_oqs}
Breuer, H.~P., and Petruccione, F., \emph{The theory of open quantum systems},
  Oxford University Press, 2002.

\bibitem[Pati and Braunstein(2000)]{nodel}
Pati, A.~K., and Braunstein, S.~L., \emph{Nature}, \textbf{404}, 164 (2000).

\bibitem[Sarovar and Milburn(2004)]{SM04}
Sarovar, M., and Milburn, G.~J., \emph{quant-ph/0501038} (2004).

\end{thebibliography}

\end{document}